\documentclass[conference,a4paper]{IEEEtran}

 \makeatletter
 \def\ps@headings{%
 \def\@oddhead{\mbox{}\scriptsize\rightmark \hfil \thepage}%
 \def\@evenhead{\scriptsize\thepage \hfil \leftmark\mbox{}}%
 \def\@oddfoot{}%
 \def\@evenfoot{}}
 \makeatother
\usepackage{graphicx}
\usepackage{amssymb}
\usepackage{algorithm}
\usepackage{algorithmic}

\newcommand{\mynotex}[1]{}

\begin{document}

\title{Improving Mobile Video Streaming with Mobility Prediction and Prefetching in Integrated Cellular-WiFi Networks
}

\author{
 Vasilios A. Siris, Maria Anagnostopoulou, and Dimitris Dimopoulos  \\
Mobile Multimedia Laboratory, Department of Informatics\\
Athens University of Economics and Business, Greece\\
}


\maketitle

\begin{abstract}
We present and evaluate a procedure that utilizes  mobility and throughput prediction to prefetch video streaming data in integrated cellular and WiFi networks. The effective integration of such heterogeneous wireless technologies will be significant for supporting high performance and energy efficient video streaming in ubiquitous networking environments.
Our evaluation is based on trace-driven simulation considering empirical measurements and shows how various system parameters influence the performance, in terms of the number of paused video frames and the energy consumption; these parameters include the number of video streams, the mobile, WiFi, and ADSL backhaul throughput, and the number of WiFi hotspots.  Also, we assess the procedure's robustness to time and throughput variability. Finally, we present our initial prototype that implements  the proposed approach.
\end{abstract}

\section{Introduction}

A major trend in mobile networks over the last few years is the exponential increase of powerful  mobile devices, such as smartphones and tablets, with multiple heterogeneous wireless interfaces that include 3G/4G/LTE and WiFi. The proliferation of such  devices has resulted in a skyrocketing growth of mobile traffic, which in 2012 grew 70\%, becoming nearly 12-times the global Internet traffic in 2000, and is expected to grow 13-fold  from 2012 until 2017\footnote{Source: Cisco Visual Networking Index: Global Mobile Data Traffic Forecast Update, 2012-2017, Feb. 6, 2013}.
Moreover, mobile video traffic was 51\% of the total traffic by the end of 2012 and is expected to become two-thirds of the world's mobile data traffic by 2017. The increase of video traffic will further intensify the strain on cellular networks, hence reliable and efficient support for video in future networks will be paramount.



The efficient, in terms of both network resource utilization and energy consumption, support for video streaming in future mobile environments with ubiquitous access will require integration of heterogeneous wireless technologies with complementary characteristics; this includes cellular networks with wide-area coverage and WiFi hotspots with high throughput and energy efficient data transfer.
Moreover, the industry has already verified the significance of mobile data offloading:
globally, 33\% of total mobile data traffic was offloaded onto  WiFi networks or femtocells in 2012$^1$.

The goal of this paper is to  evaluate the improvements for mobile video streaming  that can be achieved by exploiting  mobility and throughput prediction to prefetch video data in local storage of  WiFi hotspots, efficiently utilizing the resources of integrated cellular and WiFi networks.
Mobility prediction can provide information on the route that a vehicle will follow and when the vehicle will reach different locations along its route. Throughput prediction allows the mobile to determine whether to use WiFi hotspots that it will encounter along its route, and whether to perform prefetching.
Although we consider mobile video data offloading  to WiFi hotspots, our results and conclusions are potentially applicable to mobile video offloading to femto or small cell networks, where the backhaul throughput is smaller than the radio interface throughput.
In summary, our contributions are the following:
\begin{itemize}
\item We propose a procedure that  exploits mobility and throughput prediction for  video data prefetching, i.e., proactive caching of video data in  WiFi hotspot caches that the vehicle will encounter along its route, \emph{during} video streaming.
\item We evaluate the proposed procedure, in terms of the improved mobile video streaming Quality of Experience (QoE) and energy consumption, considering  empirical measurements and a wide range of system parameter values, and show the procedure's robustness to time and throughput variability.
\item We present the high-level design of our initial prototype, whose main component is  a  video player client for Android devices that can stream video data from different servers during video playout.
\end{itemize}
%
%
The work in this paper is different from our previous work in \cite{Sir++12,Sir++13} that considers mobile data offloading for delay tolerant traffic, which requires transferring a file until some time threshold, and delay sensitive traffic, which requires minimizing the file transfer time; unlike these  traffic types, video streaming requires a continuous transfer of video data to avoid impact on a user's QoE, which thus requires a totally different prefetching procedure and evaluation. Moreover, in this paper we present the architecture of our Android client prototype for mobile video streaming from multiple servers.

The rest of this paper is structured as follows:
Section~\ref{sec:related} presents related work, identifying how the work in this paper differs and advances the state-of-the-art.
Section~\ref{sec:procedures} presents the procedure to use mobility and throughput prediction to prefetch video data in order to improve mobile video streaming. Section~\ref{sec:evaluation} uses trace-driven simulation and empirical measurements to assess the performance and energy efficiency achieved with prefetching, and investigate it's robustness to  time and throughput variability. Section~\ref{sec:prototype} presents our initial prototype. Finally, Section~\ref{sec:conclusions} concludes the paper identifying future research directions.

\vspace{-0.05in}
\section{Related work}
\label{sec:related}
\vspace{-0.02in}

Prior work has demonstrated bandwidth  predictability  for cellular networks  \cite{Yao++08} and  WiFi \cite{Nic++08,Pan++09}.
Bandwidth prediction for improving video streaming is investigated in  \cite{Eve++11}, and for client-side pre-buffering  in \cite{Sin++12}. Both \cite{Eve++11,Sin++12} focus on cellular networks, while we consider integrated cellular-WiFi networks and prefetching video data in WiFi hotspots.
Moreover, our goal is not to develop a new system for mobility and bandwidth prediction, but to  evaluate mobility and throughput prediction to prefetch data in order to improve mobile video streaming and  investigate how time and throughput variability influence the performance gains.

Exploiting  delay tolerance to increase   mobile data offloading to WiFi  is investigated in \cite{Bal++10}. The work of \cite{Lee++10} showed that delay tolerance up to 100~seconds  provides minimal offloading gains; however, this applies to human daily mobility, rather than vehicles.
The work in \cite{Hou++11} applies a user utility model for offloading traffic to WiFi.
Our work differs in that we focus on video streaming and exploit prefetching video data in WiFi hotpots along a vehicle's route.

Work on video streaming in heterogeneous networks that exploits cooperation between devices that communicate in an ad hoc or peer-to-peer manner is investigated in \cite{Sti++09,Ana++07,Do++11,Kel++12}.
Work on exploiting multiple heterogeneous wireless interfaces is investigated in \cite{Rod++04,Tsa++09}.
Unlike the above, our work focuses on how a single mobile device can exploit  cellular and WiFi hotspot networks along its route, by using mobile and throughput prediction to prefetch video stream data in local caches of hotspots that the mobile will encounter.


The feasibility of using prediction together with prefetching is investigated in \cite{Des++09}, which  develops  a prefetching protocol (based on HTTP range requests), but does not propose or evaluate specific prefetching algorithms. In this paper we propose a procedure for video stream prefetching, and evaluate its performance and robustness against time and throughput  variations.
Prefetching to improve the performance of video file delivery in cellular femtocell networks is investigated in \cite{Gol++12}, and to reduce the peak load of mobile networks by offloading traffic to WiFi hotspots \cite{Mal++12}. Our work differs in that we consider prefetching in WiFi hotpspots along a vehicle's route to improve video streaming.



\mynotex{ Prior/related work on:
\begin{itemize}
\item Offloading of delay tolerant traffic
\item Wifi throughput prediction
\item Mobile throughput prediction, e.g. for streaming multimedia traffic.
\item Prefetching. In space-domain versus time-domain
\item mobile-WiFi handover not the focus of the paper.
\item also how prefetching can be implemented (http get) is discussed in \cite{Des++09}.
\end{itemize}
}

\mynotex{Additional notes:
\begin{itemize}
\item In this paper we investigate prediction in the time-domain, namely a nodes location in some future time. We can also exploit mobility prediction in the space-domain, i.e. to consider multiple future network attachment point where a node can move to, and prefetch data not in one location but in multiple possible future locations.
\end{itemize}

}

\section{Mobility \& throughput prediction for prefetching}
\label{sec:procedures}

In this section we present a procedure that uses mobility and throughput prediction for prefetching data in order  to improve mobile video streaming in integrated cellular-WiFi networks. Mobility prediction provides knowledge of how many WiFi hotspots a node (vehicle) will encounter, when they will be encountered, and for how long the node will be in each hotspot's range. In addition to this mobility information, we assume that information on the estimated throughput  in the WiFi hotspots and the cellular network, at different positions along the vehicles's route, is also available; for the former, the information includes both the throughput for transferring data from a remote location, e.g., through an ADSL backhaul, and the throughput for transferring data from a local cache.

Prefetching can provide gains when the throughput of transferring data from a local cache in the WiFi hotspot is higher than the throughput for transferring data from its original remote server location. This occurs when the backhaul link connecting the hotspot to the Internet has low capacity (e.g., in the case of an ADSL backhaul) or when it is congested; this is likely to become more common as the use of the IEEE 802.11n standard increases.

The  procedure to exploit mobility and throughput prediction for prefetching is shown in Algorithm~\ref{alg:streaming}.
The procedure defines the mobile's actions  when it exits a WiFi hotspot, hence has only mobile access (Line~\ref{line:mobile}), and when it enters a WiFi hotspot (Line~\ref{line:wifi}).
Mobility and throughput prediction allows the mobile to determine when it will encounter the next WiFi hotspot that has higher throughput than the cellular network's throughput. From the time to reach the next hotspot and the average video buffer playout rate, the mobile can estimate the position that the video stream is expected to reach (offset) when it arrives at the next WiFi hotspot (Line~\ref{line:offset}). In our evaluation, the average video playout rate is computing based on an EWMA\footnote{The EWMA uses weight $0.1$ for the rate in the current frame, but our experiments have shown that the results are not very sensitive to this value.} (Exponential Weighted Moving Average).
The mobile instructs a local cache in the WiFi hotpot to start caching the video stream starting from the estimated offset (Line~\ref{line:cache}).


\newcommand{\swifi}{\mbox{{\scriptsize WiFi}}}
\newcommand{\smobile}{\mbox{{\scriptsize mobile}}}
\newcommand{\sthres}{\mbox{{\scriptsize thres}}}
\newcommand{\smin}{\mbox{{\scriptsize min}}}
\newcommand{\smax}{\mbox{{\scriptsize max}}}
\newcommand{\scache}{\mbox{{\scriptsize cache}}}
\newcommand{\twifi}{\mbox{{\tiny WiFi}}}
\newcommand{\tmobile}{\mbox{{\tiny mobile}}}
\newcommand{\tthres}{\mbox{{\tiny thres}}}
\newcommand{\tmin}{\mbox{{\tiny min}}}
\newcommand{\tmax}{\mbox{{\tiny max}}}
\newcommand{\tcache}{\mbox{{\tiny cache}}}

\newcommand{\tadslno}{\mbox{{\tiny adsl}}}
\newcommand{\tbckhl}{\mbox{{\tiny bckhl}}}

\newcommand{\tnext}{\mbox{\emph {\tiny next}}}
\newcommand{\Offset}{\mbox{\emph {Offset}}}

\renewcommand{\algorithmiccomment}[1]{/* #1 */}

\algsetup{linenosize=\small}

\begin{algorithm}
\caption{Using mobility and throughput prediction to prefetch video data}
\begin{algorithmic}[1]
\label{alg:streaming}
{\scriptsize
\STATE \textbf{Variables:}
\STATE $R$: average video buffer playout rate
\STATE $T_{{\mbox{\tiny next WiFi}}}$: average time until node enters range of next WiFi
\STATE $\Offset$: estimated position in video stream when node enters next WiFi hotspot
\STATE \textbf{Algorithm:}
\IF { node exits WiFi hotspot } \label{line:mobile}
\STATE $\Offset \leftarrow  R \cdot T_{ {\mbox{\tiny  next WiFi}}}$ \label{line:offset}
\STATE Start caching video stream in next WiFi starting from $\Offset$ \label{line:cache}
\ELSIF  { node enters WiFi hotspot } \label{line:wifi}
\STATE Transfer video data that has not been received up to $\Offset$ from original  location \label{line:adsl_before}
\STATE Transfer video data from  local cache
\STATE Use remaining time in WiFi hotspot to transfer video data from original location \label{line:adsl_after}
\ENDIF
\label{line:1end}
\\ }
\end{algorithmic}
\end{algorithm}

When the node enters a WiFi hotspot, it might be missing some portion of the video stream up to the offset from which  data was cached in the hotspot; this can occur if, due to time variations, the node reaches the WiFi hotspot later than the time it had initially estimated.
In this case, the missing data needs to be transferred from the video's original remote location (Line~\ref{line:adsl_before}), through the hotspot's backhaul link. Also,  the amount of data cached in the WiFi hotspot can be smaller than the amount the node could download within the time it is in the   hotspot's range. In this  case, the node uses its remaining time in the WiFi hotspot to transfer data, as above, from the video's original location (Line~\ref{line:adsl_after}).

The procedure in Algorithm~\ref{alg:streaming} is implemented in each mobile node, e.g. in a video player client as in our prototype implementation,  Section~\ref{sec:prototype}. The scalability of the system and the accuracy of mobility and throughput prediction depends on the underlying crowdsourcing system used to collect real-time mobility and throughput information. The investigation of the scalability and prediction accuracy of the crowdsourcing system is outside the scope of this paper. However, in the next section we investigate how time and throughput variability affects the performance of the proposed prefetching procedure.

\section {Trace-driven evaluation}
\label{sec:evaluation}

\begin{figure}[t]
\centering
\includegraphics[width=3.4in]{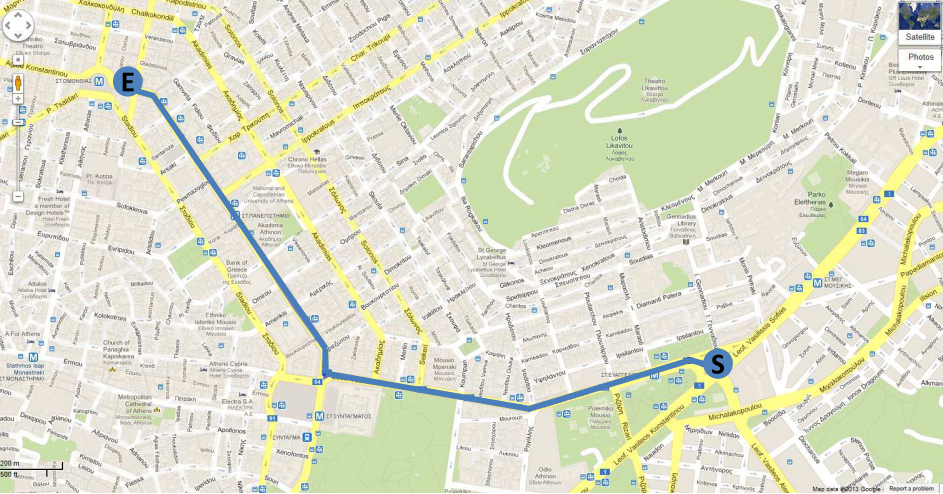}
\vspace{-0.05in}
\caption{\small{Route considered in the evaluation, along which we embed 2, 4, and 8  WiFi hotspots.
The route's total travel time from S to E is 269~seconds.}}
\label{fig:path}
\vspace{-0.15in}
\end{figure}

Our evaluation considers empirical measurements for the throughput of the cellular network and the SNR (Signal-to-Noise Ratio) of WiFi networks along a route between two locations in the center of Athens, Greece, Figure~\ref{fig:path}. Along the route
we embed 2, 4, and 8 WiFi hotspots for different scenarios investigated.
Based on the number of hotspots  we can separate the full route into segments where the moving node has either mobile  or WiFi connectivity, as shown in Table~\ref{tab:route_segments_4} for 4 hotspots (due to space limitations, we omit the corresponding tables for 2 and 8 hotspots).

\begin{table}[b]
\vspace{-0.2in}
\caption{Connectivity when 4 WiFi hotspots are embedded along the route.  }
    \label{tab:route_segments_4}
    \centering
 {\scriptsize
\begin{tabular}{|c|c|c|c|}
         \hline
         Segment  &  Access &  Time (sec) & Throughput (Mbps)\\
         \hline \hline
     1 & mobile & 0 & 4.83\\
     2 & WiFi & 18 & 16.16 (WiFi) - 6.81 (ADSL) \\
     3 & mobile & 36 & 4.22 \\
     4 & mobile & 54 & 4.58 \\
     5 & mobile & 72 & 4.58 \\
     6 & WiFi & 90 &  16.74 (WiFi) - 8.37 (ADSL) \\
     7 & mobile & 108 & 6,48 \\
     8 & mobile & 126 & 6,72 \\
     9 & mobile & 144 & 6,72 \\
     10 & WiFi & 162 & 16.74 (WiFi) - 8.37 (ADSL) \\
     11 & mobile & 180 & 4.72\\
     12 & mobile & 198 & 4.72\\
     13 & mobile & 216 & 6.51\\
     14 & WiFi & 234 & 17.23 (WiFi) - 9.46 (ADSL) \\
     15 & mobile & 252 & 5.82\\
     16 & mobile & 270 & 5.82\\
         \hline
         \end{tabular}
}
\end{table}

%

The mobile throughput  in Table~\ref{tab:route_segments_4} is the average of multiple measurements within each mobile segment.
Unlike the mobile throughput, and because the WiFi APs along the route were not open, we estimated the WiFi throughput and the throughput for downloading data over an ADSL link that would have been achieved if WiFi APs were open as follows:
We initially measured the SNR value for the various APs encountered along the route. Based on the SNR values, we estimate the throughput for downloading data stored locally at the WiFi hotspot and the throughput for downloading data over the hotspot's ADSL backhaul link using Table~\ref{tab:SNR_thrpt}; the measurements in this table were obtained empirically from public open WiFi hotspots, at various outdoor locations. Note that we are not suggesting that the mapping shown in Table~\ref{tab:SNR_thrpt} is universal. Rather, the above approach is used to obtain realistic throughput values that can be experienced in actual systems, for the specific route that we consider. Moreover,  our evaluation is not limited to the values shown Table~\ref{tab:route_segments_4}, but considers  different mobile, WiFi, and ADSL throughput values, as shown in Table~\ref{tab:values}, to investigate their impact on the  performance of the proposed video prefetching scheme.

\begin{table}[t]
\caption{Estimation of WiFi and ADSL throughput based on SNR values, using empirical measurements from open WiFi hotspots. }
    \label{tab:SNR_thrpt}
    \centering
 {\scriptsize
        \begin{tabular}{|c|c|c|}
        \hline
        SNR (dB)  &  WiFi (Mbps) &  ADSL  (Mbps)\\
        \hline \hline
    $>$ -50 & 19.90 & 15.87 \\
    -60 to -50 & 18.30 & 11.86 \\
    -70 to -60 & 17.76 & 10.13 \\
    -80 to -70 & 17.23 & 9.46 \\
    -90 to -80 & 16.74 & 8.37 \\
    $<$ -90 & 16.16 & 6.81 \\
        \hline
        \end{tabular}
}
\vspace{-0.15in}
\end{table}


The time variability determines how much the times at which the node changes access technology can differ from the empirical values shown in  Table~\ref{tab:route_segments_4}; for example, a 10\% time error means that the time the first segment (where the node has mobile access)  ends and the  second segment (where it has  WiFi access) begins is in the interval $[0.9 \cdot 18, 1.1 \cdot 18]=[16.2, 19.8]$~seconds. Note that our empirical mobility measurements indicate that under typical road traffic conditions, the timing for the various route segments can differ 10-20\%.

The throughput variability determines the throughput's deviation from its  average in Table~\ref{tab:values}; for example, a 40\%  variability means that the mobile throughput is in the interval $[0.6 \cdot M,1.4 \cdot M]$~Mbps, where $M$ is the average mobile throughput  in Table~\ref{tab:route_segments_4},  measured empirically.
We only consider the downlink direction, hence the backhaul throughput in Table~\ref{tab:values} refers to the downstream.

%
\begin{table}[b]
\vspace{-0.1in}
\caption{Parameter values. $M, W,$ and $A$, are the mobile, WiFi, and ADSL throughput for the various segments in Table~\ref{tab:route_segments_4}}
    \label{tab:values}
    \centering
 {\scriptsize
        \begin{tabular}{|c|c|}
        \hline
        Parameter  &  Values\\
        \hline \hline
   Number of streams      & 2, 3, 4 (default), 5 (for HD stream) \\
         & 8, 9, 10, 11 (default) (for SD stream) \\
    Mobile throughput  & $0.6\cdot M, 0.8\cdot M, M$ (default), $1.2 \cdot M$\\
    WiFi throughput & $0.6\cdot W, 0.8\cdot W, W$ (default), $1.2 \cdot W$ \\
    Backhaul throughput & $0.6\cdot A, 0.8\cdot A, A$ (default), $1.2 \cdot A$ \\
    Time variability & 10\% (default), 20\%, 30\%, 40\%  \\
    Throughput variability & 20\% (default), 40\%, 60\%, 80\% \\
    Number of WiFi hotspots & 2, 4 (default), 8 \\
        \hline
        \end{tabular}
        }
\end{table}


The evaluation results  are obtained from trace-driven simulation using the  parameter values in Tables~\ref{tab:route_segments_4} and \ref{tab:values}, for both HD-High Definition (Big Buck Bunny, 24 fps, avg: 1.6 Mbps, peak: 42 Mbps, peak/avg=26) and SD-Standard Definition (Football clip, 25 fps, avg: 0.6 Mbps, peak: 6.9 Mbps, peak/avg=11) video stream traces. The simulation models the video playout buffer, whose video frame data are removed when the frame is played, and is filled when data is obtained either from the original video stream location or from a local hotspot cache; in the former case the downloading rate is determined either by the mobile network or the ADSL throughput, whereas in the latter case the rate is determined by the WiFi throughput. When a frame needs to be played but the buffer does not contain the necessary data, the frame is assumed to be paused; the number of paused frames characterizes the QoE (Quality of Experience) of the video stream playout. The video playout starts after 200 frames (approximately 8 seconds) have been downloaded.
We also assume  that the WiFi interface is activated 20~seconds prior to connecting to the WiFi hotspot.
The graphs presented show the average and the 95\% confidence interval from 120~runs of each scenario, with the same parameter values.
Also,  the values in Table~\ref{tab:values} depicted as default are those that do not change in the specific evaluation scenario (graph).

\subsection{Number of paused frames}

In this section we evaluate the performance of prefetching in terms of the number of paused frames, which expresses the QoE of video streaming, and investigate how this metric is influenced by the various parameters shown in Table~\ref{tab:values}.

\begin{figure}[t]
\begin{center}
\begin{tabular}{c}

\begin{minipage}[b]{0.5\linewidth}
\centering
\hspace{-0.22in}
\includegraphics[width=1.7in] {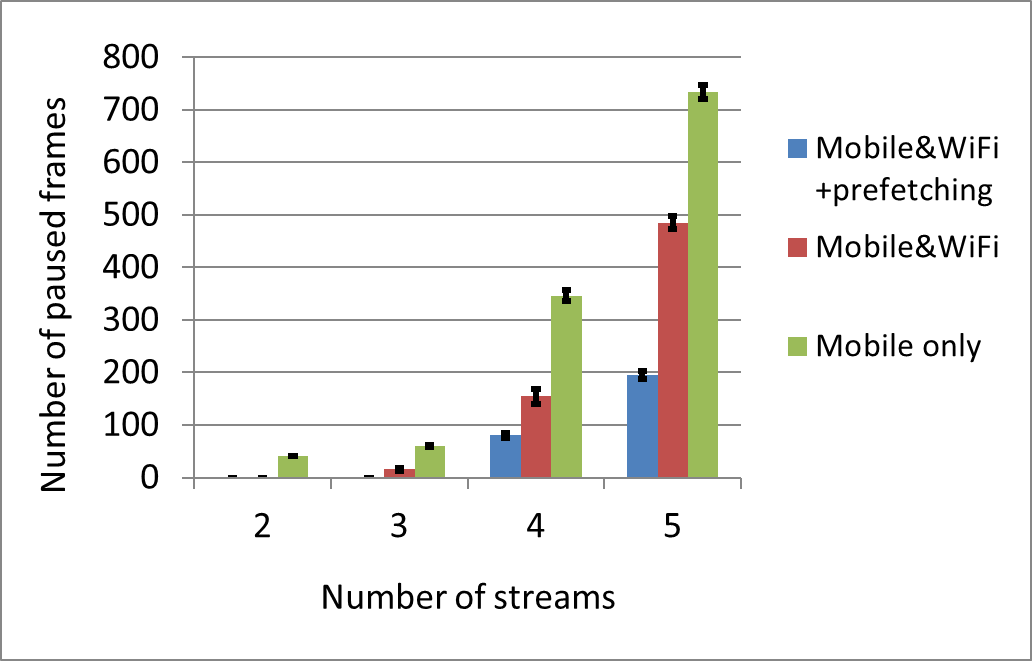}\\
{\footnotesize \small{(a) HD stream}}
\end{minipage}

\begin{minipage}[b]{0.5\linewidth}
\centering
\hspace{-0.22in}
\includegraphics[width=1.7in]{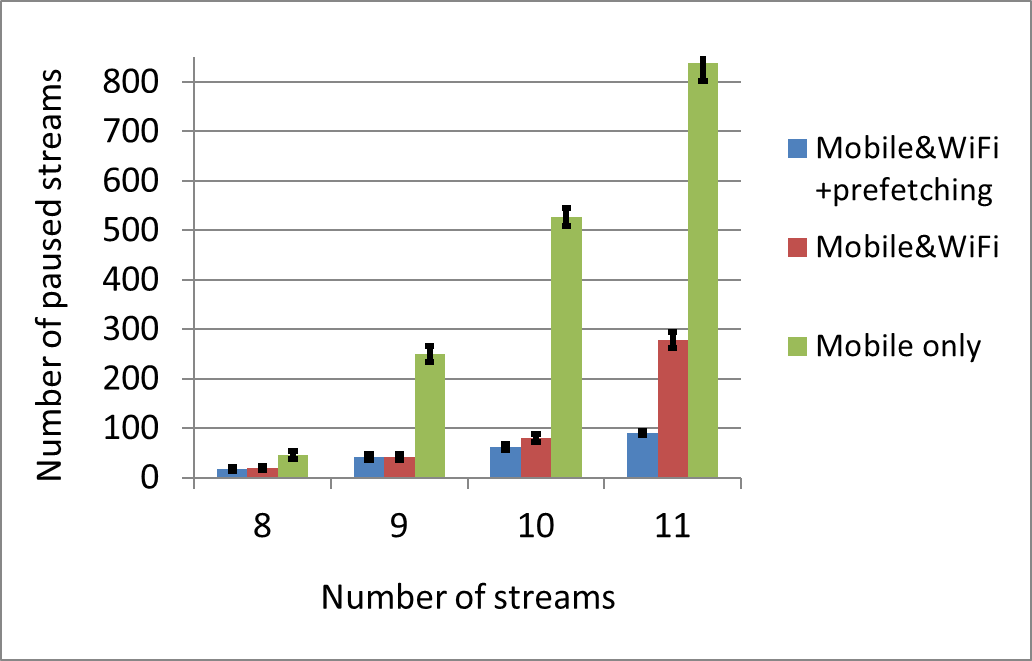}\\
{\footnotesize  \small{(b) SD stream}}
\end{minipage}\

\end{tabular}
\end{center}
\caption[]{\protect \small{Paused frames for a different number of  video streams.
}}
\label{fig:pf_streams}
\vspace{-.2 in}
\end{figure}

\medskip
\noindent
\emph{Number of video streams:} Figures~\ref{fig:pf_streams}(a) and \ref{fig:pf_streams}(b) show the number of paused frames as a function of the number of video streams, for the HD and SD video streams respectively. The results show that the gains with prefetching are high for both HD and SD video streams: For 4 HD streams, prefetching achieves  48\% and 76\% fewer paused frames compared to when prefetching is not used (i.e., the WiFi network is used opportunistically), and when only the mobile  (cellular) network is used. Moreover, note that prefetching can support 3 HD streams without paused frames, whereas WiFi without prefetching has 15 paused frames and using only the mobile network results in 60 paused frames.

For 11 SD streams, Figure~\ref{fig:pf_streams}(b) shows that prefetching achieves 68\% and 98\% fewer paused frames compared to when prefetching is not used and when only the mobile network is used, respectively. As expected, for a small number of streams, i.e., for a small  network load, the gains with prefetching are smaller.

\begin{figure}[b]
\vspace{-0.15in}
\begin{center}
\begin{tabular}{c}

\begin{minipage}[b]{0.5\linewidth}
\centering
\hspace{-0.22in}
\includegraphics[width=1.7in] {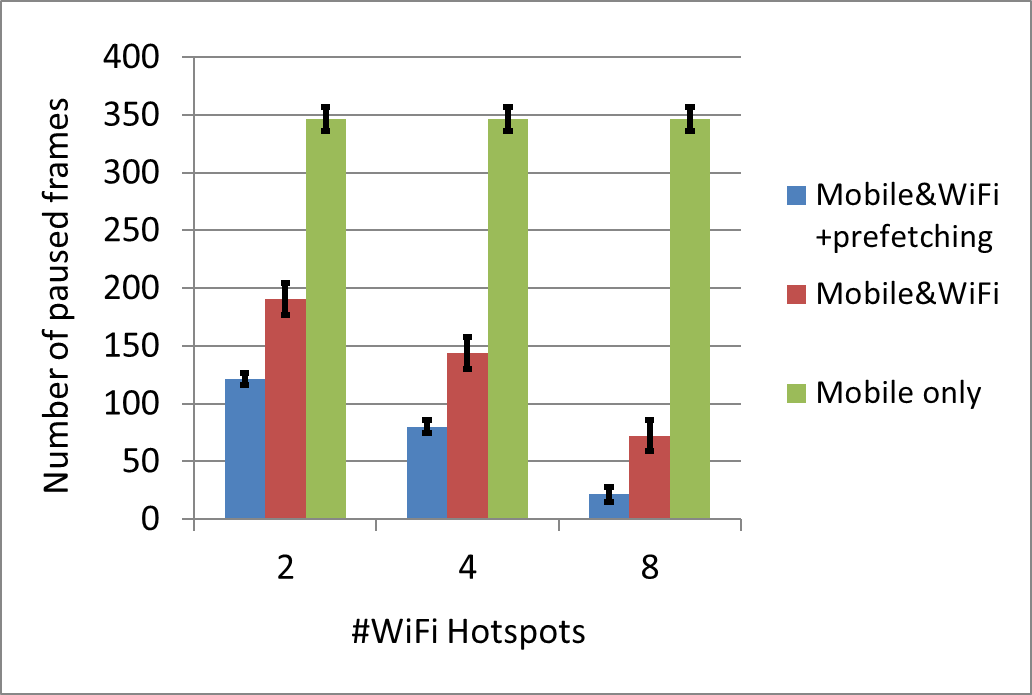}\\
{\footnotesize \small{(a) HD stream}}
\end{minipage}

\begin{minipage}[b]{0.5\linewidth}
\centering
\hspace{-0.22in}
\includegraphics[width=1.7in]{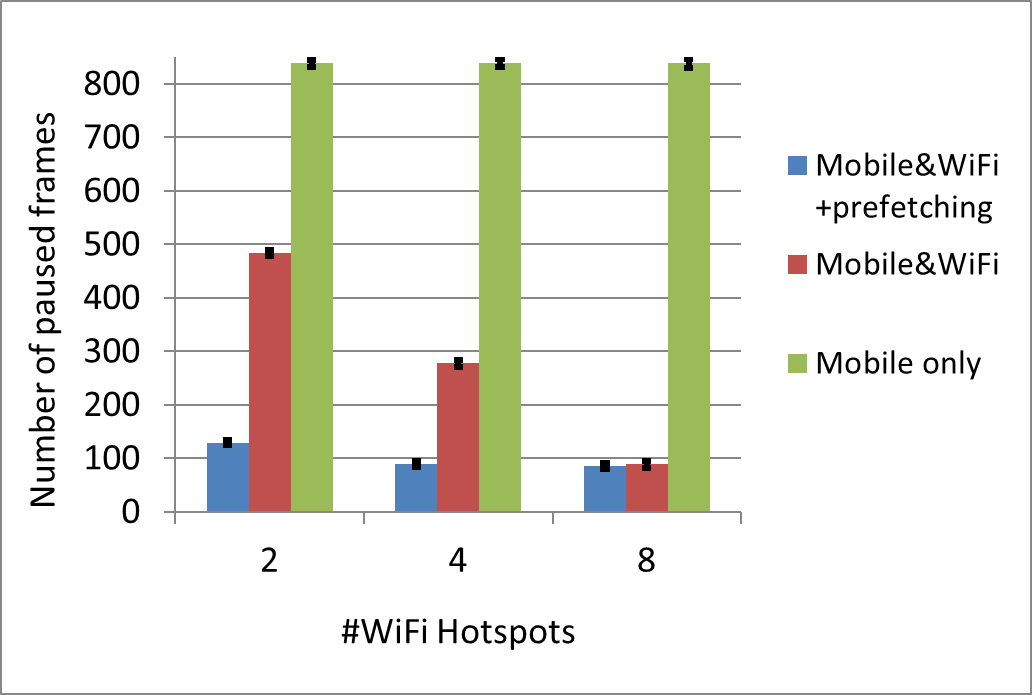}\\
{\footnotesize  \small{(b) SD stream}}
\end{minipage}\

\end{tabular}
\end{center}
\caption[]{\protect \small{Paused frames for a different number of  hotspots.
}}
\label{fig:pf_hotspots}
\end{figure}

\medskip
\noindent
\emph{Number of WiFi hotspots:} Figures~\ref{fig:pf_hotspots}(a) and \ref{fig:pf_hotspots}(b) show  that the performance for the two mobile \& WiFi schemes improves with more hotspots. For the HD stream, Figure~\ref{fig:pf_hotspots}(a), prefetching achieves more than 36\% fewer paused frames compared to when prefetching is not used.
For the SD stream, Figure~\ref{fig:pf_hotspots}(b), prefetching achieves a significantly fewer paused frames (more than 67\% fewer) for 2 and 4 hotspots. On the other hand, for 8 hotspots, prefetching does not provide gains; this happens because the remaining frame pauses occur during the initial mobile segment, which cannot be avoided with prefetching.

\begin{figure}[t]
\begin{center}
\begin{tabular}{c}

\begin{minipage}[b]{0.5\linewidth}
\centering
\hspace{-0.22in}
\includegraphics[width=1.7in] {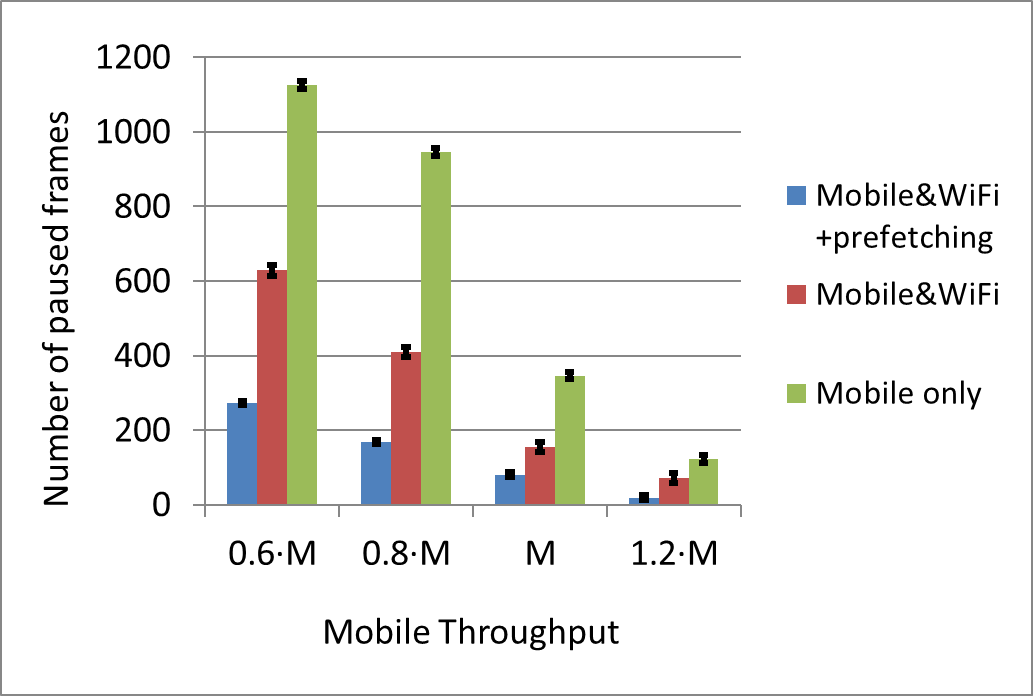}\\
{\footnotesize \small{(a) Mobile throughput}}
\end{minipage}
\begin{minipage}[b]{0.5\linewidth}
\centering
\hspace{-0.22in}
\includegraphics[width=1.7in]{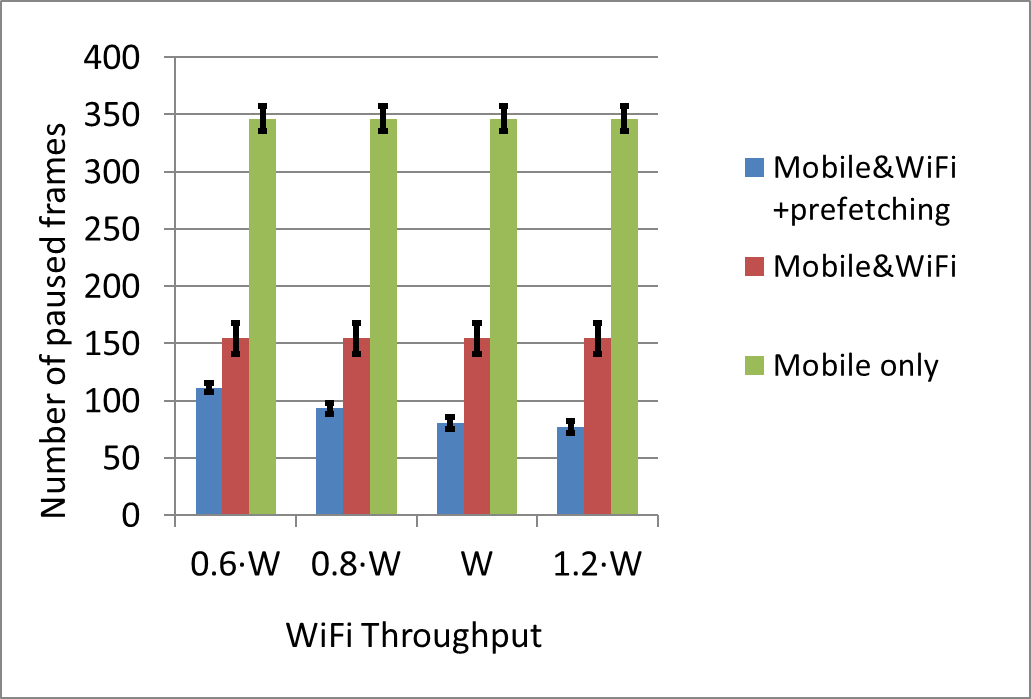}\\
{\footnotesize \small{(b) WiFi throughput}}
\end{minipage} \\ $\,$ \\
\begin{minipage}[b]{0.5\linewidth}
\centering
\hspace{-0.22in}
\includegraphics[width=1.7in]{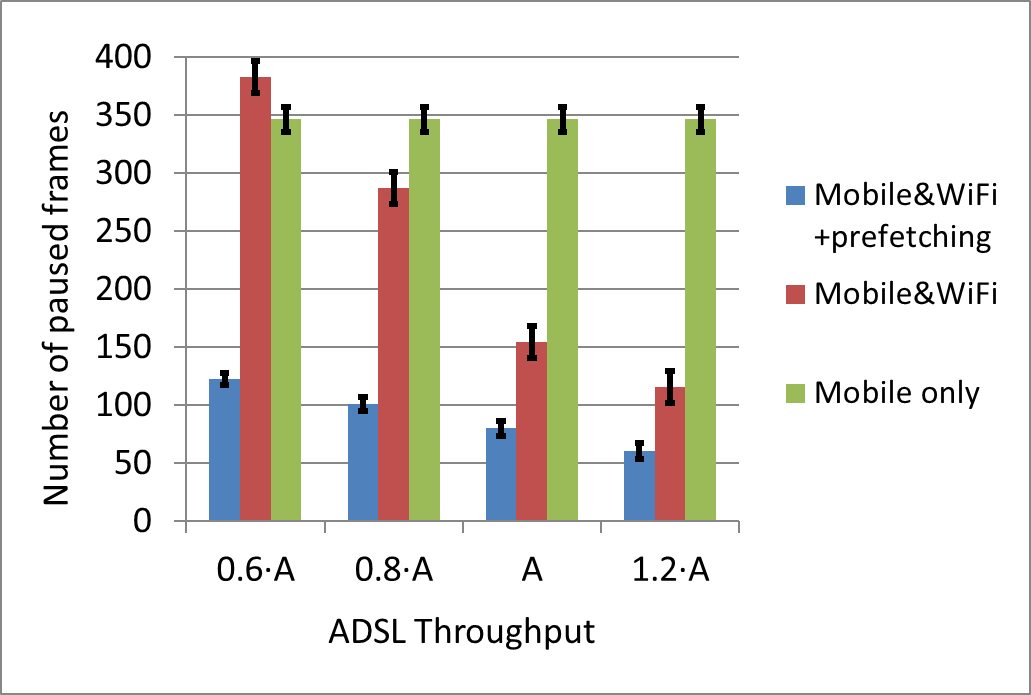}\\
{\footnotesize \small{(c) ADSL throughput}}
\end{minipage}
\end{tabular}
\end{center}
\caption[]{\protect \small{Paused frames for different mobile, WiFi, and ADSL throughput. HD  stream.
}}
\label{fig:hd_thr}
\vspace{-0.15in}
\end{figure}

\medskip
\noindent
\emph{Mobile, WiFi, and ADSL backhaul throughput:} Figure~\ref{fig:hd_thr}(a)
shows the number of paused frames for different values of the mobile throughput.
Observe that the performance for all schemes improves as the mobile throughput increases. Moreover, the performance of prefetching is significantly better compared to the other two schemes: the number of paused frames is more than 48\% lower compared to the case where prefetching is not used and more than 75\% lower compared to the case where only the mobile network is used; indeed, for mobile throughput $1.2 \cdot M$, where $M$ is the mobile throughput shown in Table~\ref{tab:route_segments_4}, the number of paused frames are less than 20, which are more than 70\% smaller than the number when WiFi is used without prefetching, and more than 84\% smaller than when only the mobile network is used.

Figure~\ref{fig:hd_thr}(b) shows the number of paused frames for different values of the WiFi throughput. As expected,  the performance with prefetching improves when the WiFi throughput increases, while the performance for mobile \& WiFi when prefetching is not affected by changes of the WiFi throughput, since the ADSL backhaul is smaller than the WiFi throughput hence limits the downloading rate.

Figure~\ref{fig:hd_thr}(c) shows the  number of paused frames for different values of the ADSL backhaul throughput. Observe that the performance for both mobile \& WiFi schemes improves as the ADSL throughput increases; however, the performance when prefetching is not used is influenced most. Indeed, when the ADSL throughput is lower than the mobile throughput, the performance when WiFi is used can be worst than when only the mobile network is used. Moreover, the performance gains of prefetching are reduced when the ADSL backhaul throughput increases, which is expected since prefetching has gains when the difference between the WiFi and the ADSL backhaul throughput is higher.

\begin{figure}[b]
\vspace{-.15 in}
\begin{center}
\begin{tabular}{c}

\begin{minipage}[b]{0.5\linewidth}
\centering
\hspace{-0.22in}
\includegraphics[width=1.7in] {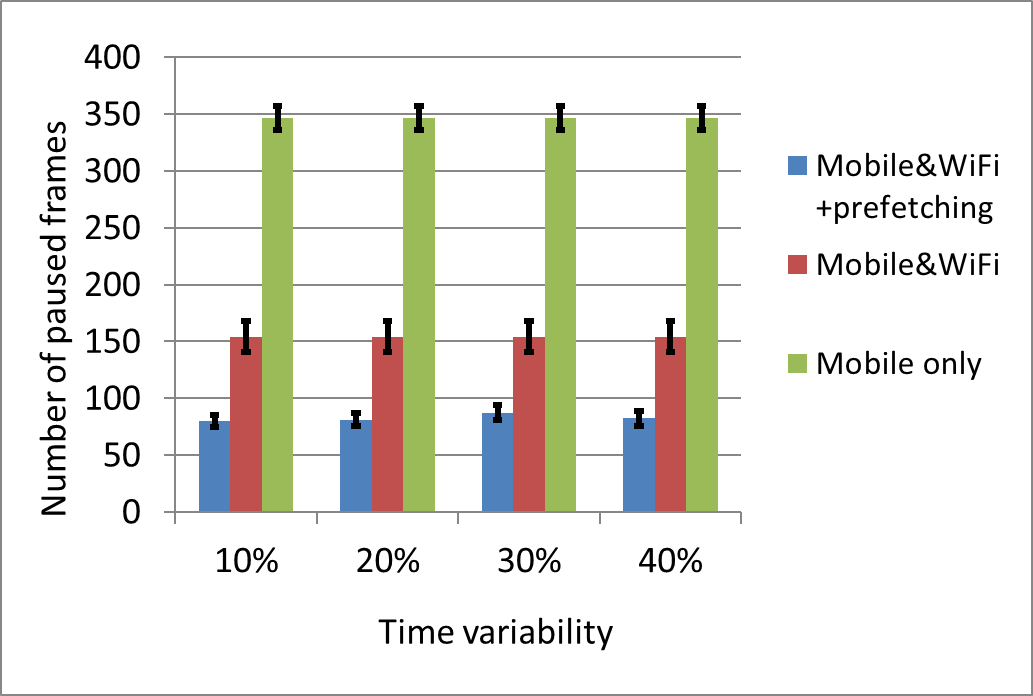}\\
{\footnotesize \small{(a) Time variability}}
\end{minipage}

\begin{minipage}[b]{0.5\linewidth}
\centering
\hspace{-0.22in}
\includegraphics[width=1.7in]{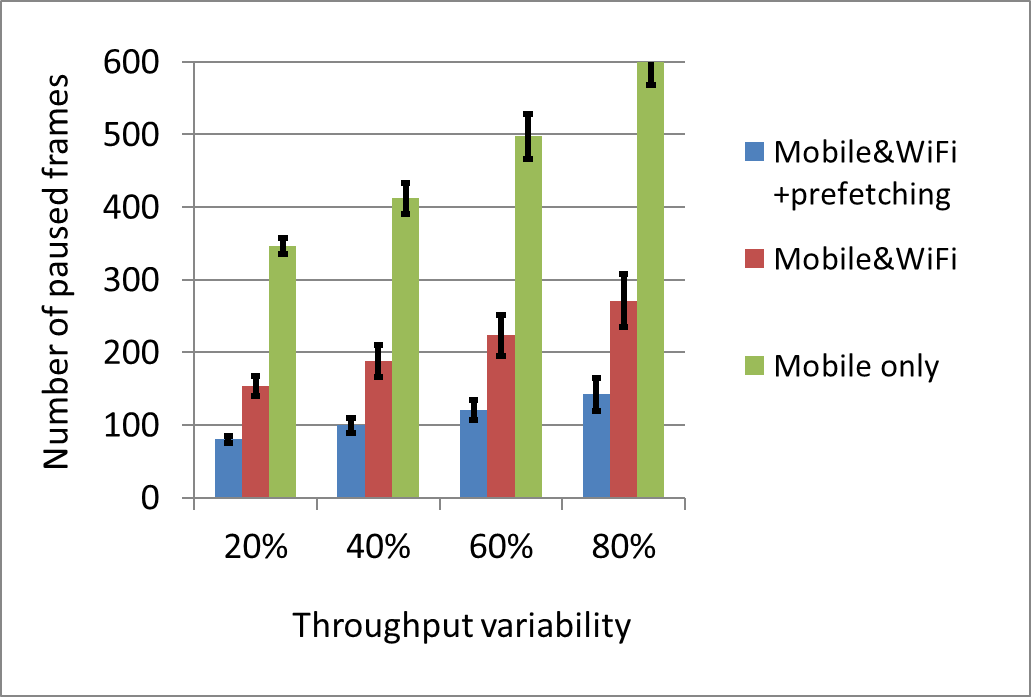}\\
{\footnotesize  \small{(b) Throughput variability}}
\end{minipage}\

\end{tabular}
\end{center}
\caption[]{\protect \small{Paused frames for different time/throughput variability. HD video stream.
}}
\label{fig:var}
\end{figure}

\medskip
\noindent
\emph{Time variability:} Figure~\ref{fig:var}(a) shows that the number of paused frames for prefetching  is not noticeably influenced by the time variability. This can be explained by the following: If a WiFi hotspot is reached later than estimated, some video data that was downloaded to a local hotspot cache  would have already been received by the mobile,  hence are not useful, but the longer time to reach the hotspot would allow more video data to be  prefetched at a local cache, and subsequently downloaded to the mobile node at the higher WiFi throughput. On the other hand, if a WiFi hotspot is reached earlier than estimated, then the the video data downloaded to a local hotspot cache would be for a later point in the video stream, but reaching the WiFi hotspot earlier would allow downloading  data at the available ADSL throughput, which if higher than the mobile throughput would improve the video streaming performance.

\medskip
\noindent
\emph{Throughput variability:} Figure~\ref{fig:var}(b) shows the influence of the variability of the mobile, WiFi, and ADSL backhaul throughput.
A higher throughput variability increases the number of paused frames for all three schemes. Nevertheless, prefetching can achieve more than 45\% fewer paused frames compared to the mobile \& WiFi scheme without prefetching, and more than 75\% fewer paused frames than when only the mobile network is used.

\vspace{-0.05in}
\subsection{Energy efficiency}
\vspace{-0.02in}

In this section we investigate the energy efficiency achieved with prefetching.
The energy  consumption is estimating using Table~\ref{tab:energy}, obtained from \cite{Ris++11}.

\begin{table}[t]
\caption{Energy consumption for 3G and WiFi, \cite{Ris++11}.}
    \label{tab:energy}
\vspace{-0.05in}
    \centering
 {\scriptsize
        \begin{tabular}{|c|c|c|}
        \hline
        Technology  &  Transfer (Joule/MB) & Idle (Watt)\\
        \hline \hline
   3G      & 100 & 0 \\
   WiFi    & 5 & 0.77 \\
        \hline
        \end{tabular}
        }
\vspace{-0.1in}
\end{table}

\begin{figure}[b]
\vspace{-0.1in}
\begin{center}
\begin{tabular}{c}

\begin{minipage}[b]{0.5\linewidth}
\centering
\hspace{-0.22in}
\includegraphics[width=1.7in] {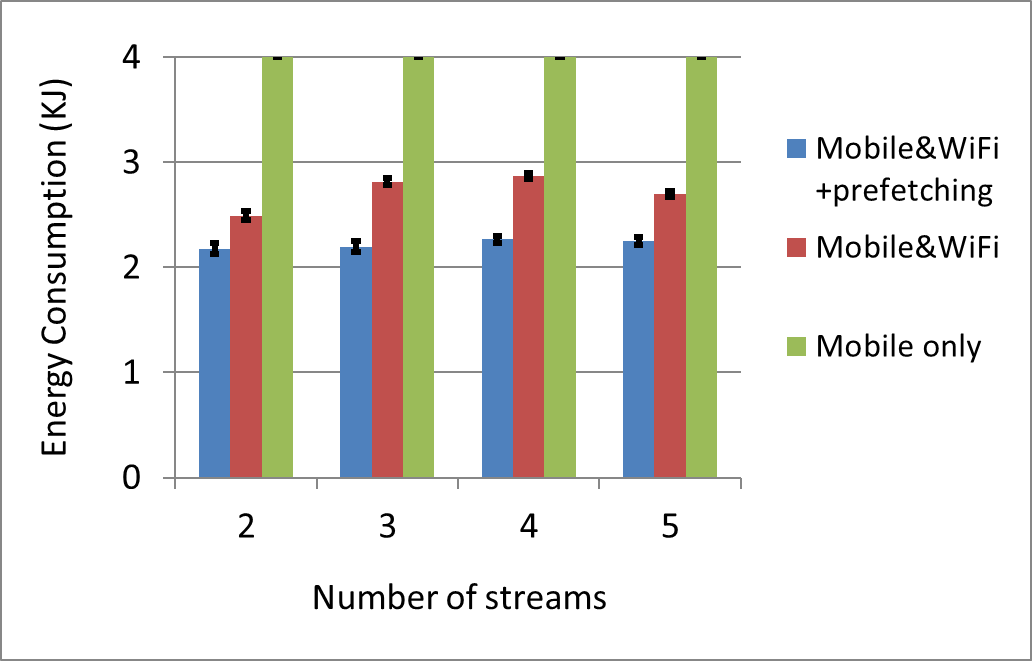}\\
{\footnotesize \small{(a) HD stream}}
\end{minipage}

\begin{minipage}[b]{0.5\linewidth}
\centering
\hspace{-0.22in}
\includegraphics[width=1.7in]{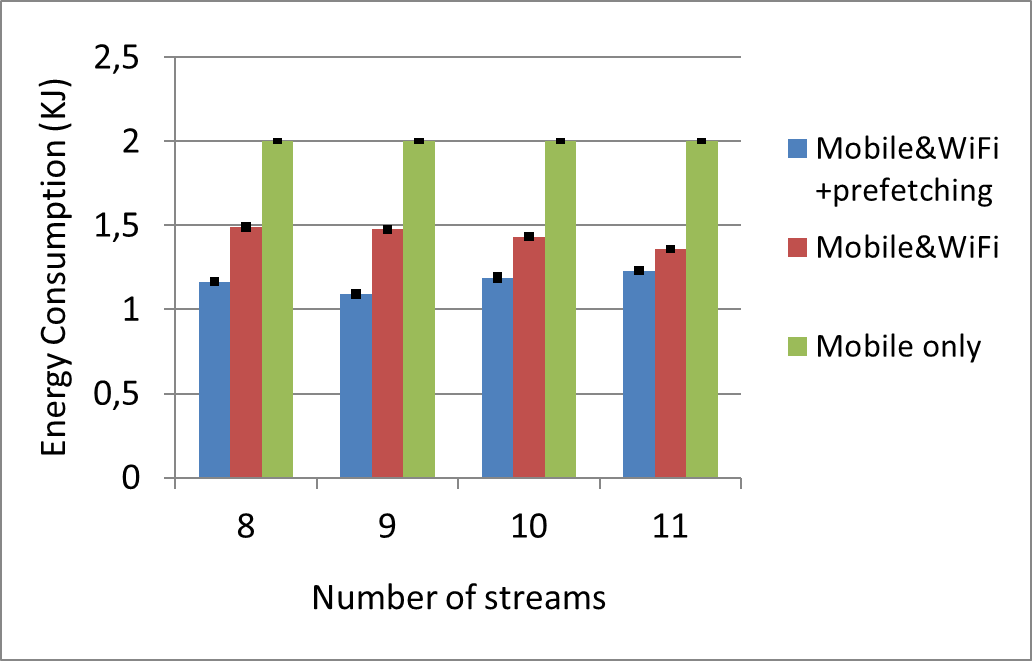}\\
{\footnotesize  \small{(b) SD stream}}
\end{minipage}\

\end{tabular}
\end{center}
\caption[]{\protect \small{Energy consumption as a function of the number of video streams.
}}
\label{fig:energy_streams}
\end{figure}

\begin{figure}[b]
\begin{center}
\begin{tabular}{c}

\begin{minipage}[b]{0.5\linewidth}
\centering
\hspace{-0.22in}
\includegraphics[width=1.7in] {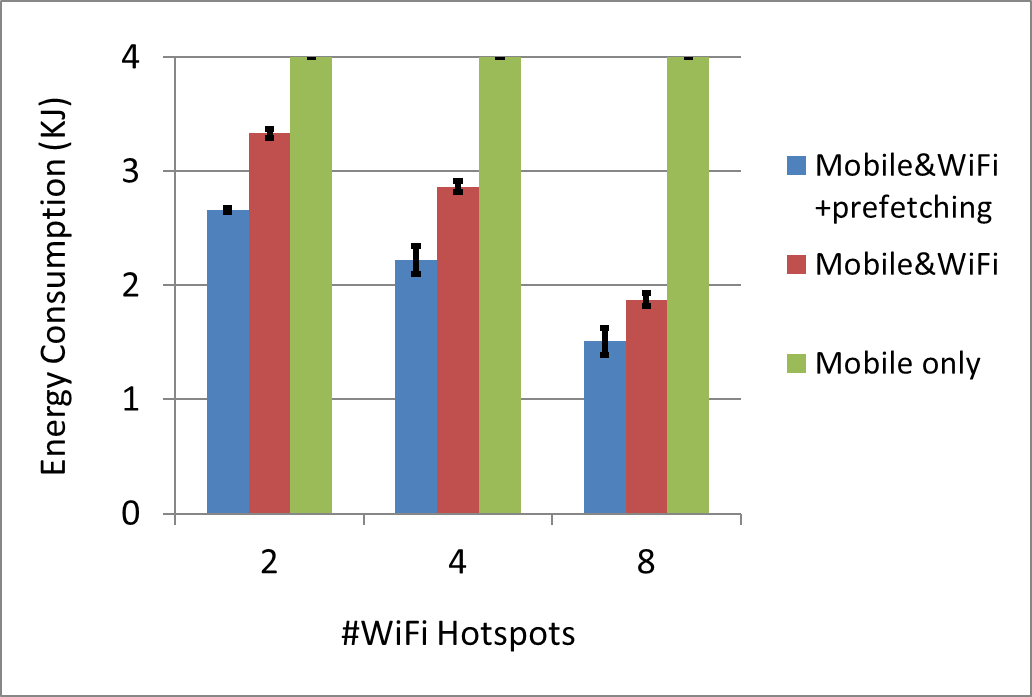}\\
{\footnotesize \small{(a) HD stream}}
\end{minipage}

\begin{minipage}[b]{0.5\linewidth}
\centering
\hspace{-0.22in}
\includegraphics[width=1.7in]{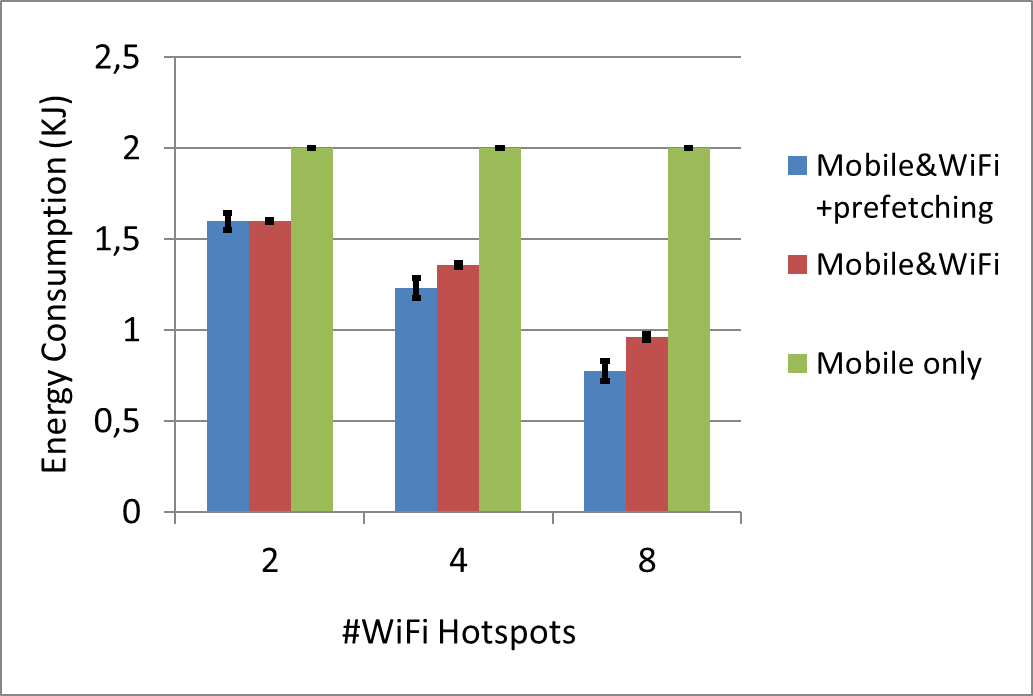}\\
{\footnotesize  \small{(b) SD stream}}
\end{minipage}\

\end{tabular}
\end{center}
\caption[]{\protect \small{Energy consumption for a different number of WiFi hotspots.
}}
\label{fig:energy_hotspots}
\end{figure}

Figures~\ref{fig:energy_streams}(a) and \ref{fig:energy_streams}(b) show the energy consumption of the investigated schemes for a different number of video streams, for HD and SD streams respectively.
Figures~\ref{fig:energy_hotspots}(a) and \ref{fig:energy_hotspots}(b) show the energy consumption of the investigated schemes for a different number WiFi hotspots, for HD and SD streams respectively. Observe that prefetching
achieves lower energy consumption compared to the case where WiFi is used without prefetching, which in turn achieves lower energy consumption compared to the case where only the mobile network is used. Nevertheless, the gains in terms of energy efficiency are comparatively lower than the gains of increased performance, in terms of a fewer number of paused frames, Figures~\ref{fig:pf_streams} and \ref{fig:pf_hotspots}. This is due to the significantly higher power consumption per transferred data of the mobile network compared to WiFi, Table~\ref{tab:energy}.

\section{Prototype implementation}
\label{sec:prototype}

\begin{figure}[t]
\centering
\includegraphics[width=3in]{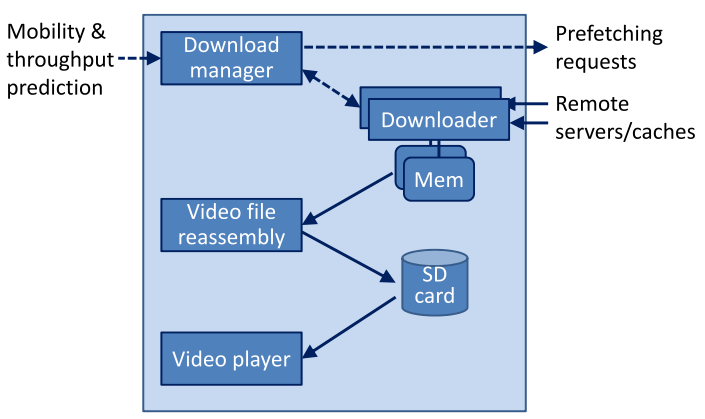}
\caption{\small{High-level design of  Android client prototype. The dashed lines represent exchange of control information, whereas the solid lines represent transfer of video data.}}
\label{fig:prototype}
\end{figure}

The high-level design of the Android client for streaming video data during video playout from multiple servers, either the original server storing the video stream or a cache, is shown in Figure~\ref{fig:prototype}.
The download manager obtains mobility and throughput prediction information, based on which it  instructs a local cache in the WiFi hotspot that the mobile client will encounter next to start caching video data starting at the estimated video offset, Algorithm~\ref{alg:streaming}.

In addition to the above functionality, the download manager, utilizing  mobility information, decides  when to turn on the WiFi interface to obtain video data from the local cache of a WiFi hotspot.
The downloader module is responsible for actually downloading video data, from the original server or the local cache. Note that our implementation allows multiple downloader threads to operate in parallel, each downloading different parts of the video stream.
However, restrictions of the Android system currently do not allow both the WiFi and 3G interfaces to operate simultaneously.
The video file reassembly module moves to the SD card, in the correct order, the parts of the video stream that are stored in the smartphone's memory by the  downloaders. The video player module reads and plays the video stream from the SD card, using Android's Media Player.

The downloader modules obtain video stream data, either from the original video stream server or from a local cache, using TCP. Note that the client needs to know a priori the IP addresses of the local hotspot caches and the original video server. The other component of the implementation is the local cache, which receives requests from the client to prefetch the part of the video file starting from some offset and sends parts of the video file to the client, when it enters the corresponding hotspot.
Currently, the communication between the mobile client and the server/local cache and between the local cache and the server uses a proprietary protocol.
The  prototype has been tested on a Android 4.0.3 (API 15) smartphone.


\section{Conclusions and Future Work}
\label{sec:conclusions}

We have presented and evaluated a procedure that exploits mobility and throughput prediction  to prefetch video data in integrated mobile and WiFi networks, in order to improve mobile video streaming. The procedure can significantly reduce the number of paused frames (QoE), while being robust to time and throughput variability and achieve reduced energy consumption.

Our evaluation considered scenarios where the cellular network is overloaded, i.e. it's throughput is smaller that the minimum throughput required to avoid frame pauses.
In situations where the cellular network is underloaded, prefetching can still provide gains in terms of reduced energy consumption. Indeed, these gains are expected to be higher than the overloaded case, since a higher percentage of the video stream  can be offloaded to WiFi. In the underloaded case the prefetching algorithm would be different than the one presented in this paper;
future work will investigate the gains achieved with prediction and prefetching  in underloaded cellular-WiFi networks.
Other ongoing work includes evaluating the performance of our prototype implementation in a realistic setting. Also, future work includes
adopting a standard protocol, based on MPEG-DASH (Dynamic Adaptive Streaming for HTTP), for communication between the mobile client and the video servers or caches.

\mynotex{Future/ongoing work
\begin{itemize}
\item use standard for client-server communication.
\item experimental test-bed evaluation.
\end{itemize}
}





\bibliographystyle{abbrv}

{\scriptsize
\bibliography{pref}
}

\end{document}